\documentclass{iopart}   
\usepackage{graphicx}   
\usepackage{dcolumn}    
\usepackage{bm}

\begin{document}

\title{Maxwell equations and the redundant gauge degree of freedom }
\author{Chun Wa Wong}
\address{Department of Physics and Astronomy, 
University of California, Los Angeles, CA 90095-1547, USA}
\ead{cwong@physics.ucla.edu}   

\begin{abstract}
On transformation to the Fourier space $({\bf k}, \omega)$, the partial 
differential Maxwell equations simplify to algebraic equations, and the 
Helmholtz theorem of vector calculus reduces to vector algebraic 
projections. Maxwell equations and their solutions can then be separated 
readily into longitudinal and transverse components relative to the 
direction of the wave vector {\bf k}. The concepts of wave motion, causality, 
scalar and vector potentials and their gauge transformations in vacuum 
and in materials can also be discussed from an elementary perspective. 
In particular, the excessive freedom of choice associated with the gauge 
dependence of the scalar and the longitudinal vector potentials stands out 
with clarity in Fourier spaces. Since these potentials are introduced to
represent the instantaneous longitudinal electric field, the actual 
cancellation in the latter of causal contributions arising from these 
potentials separately in most velocity gauges becomes an important issue. 
This cancellation is explicitly demonstrated both in the Fourier space, and
for pedagogical reasons again in space-time. The physical origin of the 
gauge degree of freedom in the masslessness of the photon, the quantum of 
electromagnetic wave, is elucidated with the help of special relativity
and quantum mechanics.
\end{abstract}

\pacs{01.30.Rr,01.50.Zr,01.55.+b,02.30.Nw}
\maketitle

\section{Introduction}
\label{sect:Introduction}

Most textbooks discuss the basic properties of the Maxwell equations and their 
solutions in ordinary space-time $({\bf r}, t)$ \cite{Jackson99,Griffiths99}. 
Electromagnetic properties in vacuum and in materials are often nonlocal in 
space-time, and are therefore awkward to describe. It is well-known that 
many of these properties become more transparent in the Fourier space 
$({\bf k}, \omega)$ reached by Fourier transforms. We present in this paper 
a pedagogical discussion of Maxwell equations in this Fourier space where 
their relative simplicity makes them more understandable at an 
introductory level. Electromagnetism is a vast subject. The main objective 
of this paper is to elucidate the mathematical and physical origins of the 
mysterious gauge degree of freedom in the choice of electromagnetic scalar 
and vector potentials. It had taken a century of research, from the 
Coulomb force law of 1785 through the 1835 vector potential of Gauss to the 
Lorenz gauge of 1867, to establish the nonuniqueness of two of these 
potential components \cite{Jackson99,Jackson01}. Gauge transformations are
discussed in almost all textbooks on electromagnetism, but the subject 
still appears mysterious to many students. 

We begin, in section \ref{sect:LTMaxwellEqs}, with the space-time Maxwell 
equations in materials in SI units. The Fourier electromagnetic fields are 
then introduced, and separated into longitudinal/transverse (L/T) 
components relative to the wave vector {\bf k}. The two Maxwell equations 
involving divergences are found to define longitudinal field components 
that in vacuum are instantaneous, acting at a distance. The two curl 
equations also involve the transverse fields. 

Section \ref{sect:WaveMotion} describes electromagnetic wave motion in 
linear isotropic materials. Electromagnetic waves involve transverse fields 
only. The electromagnetic wave equation contains a term dependent on both 
the time differential operator and a light speed $v \leq c$ in the material 
($c$ being the light speed in vacuum). Both features are needed for the
transverse fields to describe events happening {\it causally} at the light 
speed of the material. 

In section \ref{sect:SVpotentials}, the transverse part of a vector potential 
is used to describe the magnetic induction field. The longitudinal part of 
the vector potential is used together with a scalar potential to describe 
the instantaneous longitudinal electric field when either one of them alone 
can do the job. It is traditional to use a combination of both potentials. 
The permissible use of different combinations gives rise to an extra ``gauge'' 
degree of freedom. Most of these gauge choices lead to potentials that are 
causal because they satisfy wave equations where signals are propagated 
with an arbitrary gauge velocity. We show explicitly that the resulting
arbitrary gauge causal contributions from the two potentials cancel exactly
to give a longitudinal electric field that is instantaneous, acting at a 
distance, as dictated by the Gauss law of electrostatics.

In section \ref{sect:Comparison}, our results are compared with the 
space-time results of Brill and Goodman \cite{Brill67} and of Yang 
\cite{Yang05}. Although the technical analyses are similar, we find that 
the longitudinal electric field is instantaneous, not retarded, even though
it is the sum of two retarded terms. The meaning of action at a distance 
is reviewed from the modern perspective of quantum mechanics and quantum 
field theory.

In teaching electromagnetism, it is important to present the cancellation 
of the arbitrary gauge causality in the physical instantaneous 
longitudinal electric field in an intuitive and appealing way. We achieve
this objective in section \ref{sect:Eparallel} by finding a 
causality-canceling combination in space-time. Its gauge-dependent and 
causal parts associated with the scalar and longitudinal vector potentials 
are then found in {\bf k} space. A final transformation back to space-time 
shows why why the simple expressions in {\bf k} space look so complicated 
in space-time. We also use the folding theorem of Fourier transforms as 
a qualitative test for instantaneity in time and 
localization in space of physical phenomena.

In section \ref{sect:GaugeX}, the relations or transformations between 
different gauge choices are described. Even after a gauge is chosen, the 
scalar and longitudinal vector potentials can still vary by amounts
proportional to the infinitely many gauge functions that are solutions
of a certain homogeneous wave equation.  

Finally, in section \ref{sect:GaugeOrigin}, we show that the gauge degree 
of freedom appears because light speed $c$ in vacuum has the same 
value in all acceptable inertial frames called Lorentz frames. There is 
then always a longitudinal direction along which the magnetic induction 
vanishes, and the electric field component is instantaneous, acting at a 
distance. In the quantum description of light as a particle called the 
photon, this constancy of light speed comes from the masslessness of the 
photon. A massive photon, if it existed, would have a well-defined and 
dynamically meaningful longitudinal vector potential. As the photon loses 
its mass, its longitudinal vector potential decouples from the two 
transverse components, and becomes the redundant appendage called 
the gauge degree of freedom.

\section{Longitudinal and transverse Maxwell equations in materials}
\label{sect:LTMaxwellEqs}

The Maxwell equations for electromagnetic fields in ordinary materials in 
space-time $({\bf r}, t)$ in SI units are, in the notation of Jackson 
(\cite{Jackson99}, p. 248):
\begin{eqnarray}
\bm {\nabla \cdot} {\bf B} &=& 0, 
\label{monopole} \\
\bm {\nabla \times} {\bf E} &=&  - \partial_t {\bf B}, 
\label{faraday} \\
\bm {\nabla \cdot} {\bf D} 
= \bm {\nabla \cdot} (\epsilon_0 {\bf E} + {\bf P}) &=& \rho, 
\label{gauss} \\
\bm {\nabla} \bm{\times} {\bf H} = \bm {\nabla} \times 
\left (\frac{1}{\mu_0}{\bf B} - {\bf M}\right ) 
&=& {\bf J} + \partial_t {\bf D}.
\label{maxwell}
\end{eqnarray}
Here $\partial_t \equiv \partial/\partial t$, while $\rho = \rho({\bf r},t),  
{\bf J}$ are the external charge and current densities. The vector fields
${\bf E} = {\bf E}({\bf r},t), {\bf D}, {\bf P}, {\bf B}, {\bf H}, {\bf M}$
are assumed to have well-defined second space-time derivatives so that wave 
equations can be constructed. The polarization {\bf P} and the magnetization 
{\bf M} describe the responses of the material medium to the presence of
external charges and currents. Note that the first two equations hold only
in the absence of magnetic charges. When magnetic charges are present, an 
additional term $\rho_m$ and $-{\bf J}_m$ should appear on the right of 
(\ref{monopole}) and (\ref{faraday}), respectively. 

These fields and their first and second derivatives in space-time are 
assumed to have the Fourier representations
\begin{eqnarray}
{\bf E}({\bf r}, t) = \int_{-\infty}^\infty \frac{d\omega}{2\pi} 
\int \frac{d^3k}{(2\pi )^3} e^{i{\bf k}\bm{\cdot}{\bf r} - i\omega t} 
\tilde{\bf E}({\bf k}, \omega), 
\label{FR-E} \\
\bm {\nabla \cdot}{\bf E}({\bf r}, t) 
= \int_{-\infty}^\infty \frac{d\omega}{2\pi} 
\int \frac{d^3k}{(2\pi )^3} e^{i{\bf k}\bm{\cdot}{\bf r} - i\omega t} 
i{\bf k}\bm{\cdot}\tilde{\bf E}({\bf k}, \omega), 
\label{FR-divE} 
\end{eqnarray}
etc. In (\ref{FR-divE}), the space-time differential operator $\bm{\nabla}$
acts directly on the Fourier basis function 
\begin{eqnarray}
\psi_{{\bf k}, \omega}({\bf r}, t) 
= e^{i{\bf k}\bm{\cdot}{\bf r} - i\omega t}.
\label{FourierBasis} 
\end{eqnarray}

The Maxwell equations in the Fourier space $({\bf k}, \omega)$ become the 
algebraic equations
\begin{eqnarray}
i{\bf k}\bm{\cdot} \tilde{\bf B} &=& 0, 
\label{monopole-kw} \\
i{\bf k}\bm{\times} \tilde{\bf E} &=&  i\omega \tilde{\bf B}, 
\label{faraday-kw} \\
i{\bf k}\bm {\cdot} \tilde{\bf D} 
= i{\bf k}\bm{\cdot} (\epsilon_0 \tilde{\bf E} + \tilde{\bf P}) &=& \tilde{\rho}, 
\label{gauss-kw} \\
i{\bf k}\bm{\times} {\bf H} = i{\bf k} \bm{\times} 
\left (\frac{1}{\mu_0}\tilde{\bf B} - \tilde{\bf M}\right ) 
&=& \tilde{\bf J} - i\omega\tilde{\bf D}.
\label{maxwell-kw}
\end{eqnarray}
These equations can be decomposed, term by term, into longitudinal 
(L or $\parallel$) and transverse (T or $\perp$) components that are 
respectively parallel and perpendicular to {\bf k}. In the rectangular 
coordinate system defined by the unit vectors
${\bf e}_1, {\bf e}_2$ and ${\bf e}_3 = {\bf e}_{\bf k} = {\bf k}/k$, the
L/T projection operators are the dyadics or tensors \cite{Schwinger98}
\begin{eqnarray}
\tilde{\bf I}_\parallel &=& {\bf e}_{\bf k}{\bf e}_{\bf k}, 
\nonumber \\
\tilde{\bf I}_\perp &=&  \tilde{\bf I} - \tilde{\bf I}_\parallel 
= {\bf e}_1{\bf e}_1 + {\bf e}_2{\bf e}_2.
\label{LTprojectors}
\end{eqnarray}
The L/T separation of a vector field $\tilde{\bf E}$ can then be realized by 
using scalar and vector products
\begin{eqnarray}
\tilde{\bf I}_\parallel \bm{\cdot} \tilde{\bf E} 
&=& {\bf e}_{\bf k}{\bf e}_{\bf k} \bm{\cdot} \tilde{\bf E} 
= \tilde{E}_\parallel {\bf e}_{\bf k} = \tilde{\bf E}_\parallel 
= \tilde{\bf E} \bm{\cdot} \tilde{\bf I}_\parallel , 
\nonumber \\
\tilde{\bf I}_\perp \bm{\cdot} \tilde{\bf E} 
&=& - {\bf e}_{\bf k} \bm{\times} ({\bf e}_{\bf k} \bm{\times} \tilde{\bf E}) 
= \tilde{\bf E}_\perp = \tilde{\bf E} \bm{\cdot} \tilde{\bf I}_\perp. 
\label{LTprojections}
\end{eqnarray}
Written as a single equation, 
\begin{eqnarray}
\tilde{\bf E} = \tilde{\bf E}_\parallel + \tilde{\bf E}_\perp
= {\bf e}_{\bf k} ({\bf e}_{\bf k} \bm{\cdot} \tilde{\bf E}) 
- {\bf e}_{\bf k} \bm{\times} ({\bf e}_{\bf k} \bm{\times} \tilde{\bf E}), 
\label{HelmholtzThm}
\end{eqnarray}
this simple algebraic L/T separation is just the Helmholtz theorem of vector 
calculus, now reduced to simple vector algebraic projections in the Fourier 
space $({\bf k}, \omega)$. We shall show in \ref{sect:HelmholtzThm} that an 
inverse Fourier transform of (\ref{HelmholtzThm}) gives the usual integral 
statement of the Helmholtz theorem in space. The scalar and vector products
\begin{eqnarray}
{\bf k} \bm{\cdot} \tilde{\bf E} = k\tilde{E}_\parallel, \quad 
{\bf k} \bm{\times} \tilde{\bf E} = {\bf k} \bm{\times} \tilde{\bf E}_\perp
\label{SCproducts}
\end{eqnarray}
in the L/T separated form appear repeatedly in the Maxwell equations.

With L/T separation, the two Maxwell divergence equations become the scalar
algebraic equations
\begin{eqnarray}
ik\tilde{B}_\parallel &=& 0, 
\label{divEq1} \\
ik \tilde{D}_\parallel 
= ik (\epsilon_0 \tilde{E}_\parallel + \tilde{P}_\parallel) &=& \tilde{\rho}. 
\label{divEq2}
\end{eqnarray}
for the $\parallel$ components. In free space, where $\tilde{P}_\parallel$ 
vanishes, the $\omega$ dependence of $\tilde{E}_\parallel$ is identical to 
that of $\tilde{\rho}$. Their time dependences in space-time are therefore
identical too. Hence $E_\parallel({\bf r},t)$ responds to the source charge 
density $\rho({\bf r},t)$ instantaneously, acting at a distance, as we
shall show with more detail in section \ref{sect:Eparallel}.

Each of the two Maxwell curl equations separates into individual T and L 
equations
\begin{eqnarray}
i{\bf k}\bm{\times} \tilde{\bf E}_\perp &=&  i\omega \tilde{\bf B}_\perp, 
\label{curlEq1T} \\
i\omega \tilde{\bf B}_\parallel &=& 0, 
\label{curlEq1L} \\
i{\bf k}\bm{\times} {\bf H}_\perp = i{\bf k} \bm{\times} 
\left (\frac{1}{\mu_0}\tilde{\bf B}_\perp - \tilde{\bf M}_\perp\right ) 
&=& \tilde{\bf J}_\perp - i\omega\tilde{\bf D}_\perp,
\label{curlEq2T}\\
\tilde{\bf J}_\parallel - i\omega\tilde{\bf D}_\parallel &=& 0.
\label{curlEq2L}
\end{eqnarray}
The result $\tilde{B_\parallel} = 0$, stating the absence of magnetic 
charges, thus appears in two separate Maxwell equations (\ref{divEq1}) and 
(\ref{curlEq1L}). Equations (\ref{divEq2}) and (\ref{curlEq2L}) taken 
together gives the continuity equation
\begin{eqnarray}
- i\omega \tilde{\rho} + ik \tilde{J}_\parallel = 0.
\label{continuityEq}
\end{eqnarray}
In space-time, ${\bf J} = {\bf v}\rho$, where {\bf v} is the velocity of the 
charge. Then the continuity equation
\begin{eqnarray}
\partial_t\rho + \bm{\nabla \cdot}({\bf v}\rho) 
= \partial_t\rho + {\bf v}\bm{\cdot \nabla}(\rho)
= \frac{d}{dt}\rho = 0
\label{continuityEqrt}
\end{eqnarray}
describes charge conservation. Note in particular that charge conservation
does not involve $\tilde{J}_\perp$. 

We should add that since $\tilde{\bf E}({\bf k}, \omega)$ is the Fourier 
transform ${\cal F}$ of ${\bf E}({\bf r}, t)$, namely
\begin{eqnarray}
\tilde{\bf E}({\bf k}, \omega) = {\cal F}\{{\bf E}\} 
= \int_{-\infty}^\infty dt 
\int d^3r e^{-i{\bf k}\bm{\cdot}{\bf r} + i\omega t} 
{\bf E}({\bf r}, t), 
\label{FT-E} 
\end{eqnarray}
the representation (\ref{FR-E}) is just its inverse Fourier transform. The 
downside of using Fourier spaces is that it is necessary to perform this Fourier 
inversion if one wants the result in space-time. So at an introductory level, 
Fourier spaces are most useful for the qualitative understanding of an 
electromagnetic concept for which the complete Fourier inversion back to 
space-time is not essential.

\section{Wave motion in linear isotropic materials}
\label{sect:WaveMotion}

In linear materials, the contributions of charges and currents induced in 
the material are included through the dielectric tensor (or dyadic) 
$\tilde{\bm{\epsilon}}$ and the permeability tensor $\tilde{\bm{\mu}}$:  
\begin{eqnarray} 
\tilde{\bf D} = \tilde{\bm{\epsilon}} \bm{\cdot} \tilde{\bf E}, 
\qquad \tilde{\bf B} = \tilde{\bm{\mu}} \bm{\cdot} \tilde{\bf H}.
\label{epsilonMu}
\end{eqnarray}
We shall restrict ourselves to isotropic materials where the only preferred 
direction is ${\bf e}_{\bf k}$. The chosen rectangular coordinate axes in 
{\bf k} space are therefore also the principal axes of these physical 
properties. Thus
\begin{eqnarray} 
\tilde{\bm{\epsilon}} = \tilde\epsilon_\parallel \tilde{\bf I}_\parallel +
\tilde\epsilon_\perp \tilde{\bf I}_\perp: \qquad 
\tilde{\bf D} = \tilde\epsilon_\parallel \tilde{\bf E}_\parallel
+ \tilde\epsilon_\perp \tilde{\bf E}_\perp, 
\label{dielectric}
\end{eqnarray}
where the dielectric functions $\tilde\epsilon_\parallel, \tilde\epsilon_\perp$ 
are scalar functions of $k, \omega$ only. To simplify subsequent formulas, 
we shall use a cruder approximation for magnetic properties:
\begin{eqnarray} 
\tilde\mu_\parallel = \tilde\mu_\perp = \tilde\mu(k,\omega):  \qquad 
\tilde{\bf B} = \tilde\mu \tilde{\bf H}.
\label{permeability}
\end{eqnarray}

For these linear isotropic materials, (\ref{divEq2}) simplifies to 
\begin{eqnarray}
\tilde{E}_\parallel = -\frac{i}{k\tilde\epsilon_\parallel}\tilde{\rho}. 
\label{Eparallel}
\end{eqnarray}
Thus $\tilde{E}_\parallel$ is completely determined in the Fourier space 
$({\bf k}, \omega)$, and can be found in space-time by calculating 
its (inverse) Fourier transform.

The two curl equations for the transverse fields are more complicated: 
\begin{eqnarray}
i{\bf k}\bm{\times} \tilde{\bf E}_\perp &=&  i\omega \tilde{\bf B}_\perp, 
\nonumber  \\
i{\bf k}\bm{\times} {\bf B}_\perp &=& 
\tilde\mu\left( \tilde{\bf J}_\perp - 
i\omega\tilde\epsilon_\perp\tilde{\bf E}_\perp \right).
\label{curlEqsLinearMaterials}
\end{eqnarray}
After operating on these equations with $i{\bf k}\bm{\times}$, we can solve
for the $\parallel$ fields. The results are
\begin{eqnarray}
\left( k^2 - \frac{\omega^2}{v^2} \right) \tilde{\bf E}_\perp
&=& i\tilde\mu\omega {\bf J}_\perp, \nonumber \\
\left( k^2 - \frac{\omega^2}{v^2} \right) \tilde{\bf B}_\perp
&=& i\tilde\mu {\bf k} \bm{\times} {\bf J}_\perp,
\label{waveEqsFS}
\end{eqnarray}
where
\begin{eqnarray}
v(k, \omega) = \frac{1}{\sqrt{\tilde\mu\tilde\epsilon_\perp}}
\label{waveVelocity} 
\end{eqnarray} 
has the dimension of a velocity. These are inhomogeneous wave equations in 
the Fourier space $({\bf k}, \omega)$, and $v$ is the wave velocity or speed
in the material for the given $k, \omega$ values.

To see that these equations describe waves, we go back to the associated 
partial differential equations in space-time with the help of (\ref{FR-divE})
under the simplifying assumption that $\tilde{\mu}$, $\tilde{\epsilon_\perp}$ 
and $v$ are independent of $k, \omega$:
\begin{eqnarray}
\left( \nabla^2 - \frac{1}{v^2}\partial^2_t\right) {\bf E}_\perp({\bf r},t)
&=& \mu\partial_t {\bf J}_\perp({\bf r},t), \nonumber \\
\left( \nabla^2 - \frac{1}{v^2}\partial^2_t\right) {\bf B}_\perp({\bf r},t)
&=&  - \mu \bm{\nabla \times} {\bf J}_\perp({\bf r},t).
\label{waveEqsSpT}
\end{eqnarray}
These partial differential equations are called wave equations, and their 
solutions, here the transverse fields, may be called wave functions. 
We shall not describe the properties of these wave equations and their 
solutions in space-time, as these properties can be found in most textbooks 
of electromagnetism. We would only point out that it is the presence on the 
left side of each partial differential equation of the term containing 
both the time derivative $\partial_t$ ($-i\omega$ in the Fourier space)
and the wave speed $v$ that makes the wave functions causal quantites
involving signals propagating with the speed $v$. Without this term,
the differential equations would have been reduced to Poisson equations.
Then ${\bf B}_\perp({\bf r},t)$ would be instantaneous with 
${\bf J}_\perp({\bf r},t)$ if $\mu$ is independent of $t$ (or $\omega$ in
the Fourier space). ${\bf E}_\perp({\bf r},t)$ too would be 
instantaneous with $\partial_t {\bf J}_\perp({\bf r},t)$, but not 
instantaneous with ${\bf J}_\perp({\bf r},t)$ itself. This is because 
$\partial_t{\bf J}_\perp({\bf r},t)$ at one time involves more than one 
time value of ${\bf J}_\perp({\bf r},t)$. A more detailed explanation will 
be given in section \ref{sect:Eparallel}.

In many textbooks, the complete ${\bf E, B}$ fields are expressed in a
causal, actually retarded, form, when it is only their transverse parts 
${\bf E}_\perp$ and ${\bf B}_\perp$ that are retarded, according to 
(\ref{waveEqsSpT}). Since (\ref{divEq1}) and (\ref{curlEq1L}) require 
${\bf B}_\parallel = 0$, it is true that the complete 
${\bf B} = {\bf B}_\perp$ is retarded. However, ${\bf E}_\parallel$ does
not vanish in general. We shall show in the next section how the correct 
noncausal nature of $\tilde{E}_\parallel$ can be recovered from certain 
intermediate quantities that are both causal and dependent on an 
arbitrarily chosen ``gauge'' velocity.

\section{Scalar and vector potentials and their gauge degree of freedom}
\label{sect:SVpotentials}

The Helmholtz theorem (\ref{HelmholtzThm}) shows that 
\begin{eqnarray}
\tilde{\bf B} = \tilde{\bf B}_\perp 
= i {\bf k} \bm{\times} \tilde{\bf A}_\perp, 
\quad {\rm where} \quad \tilde{\bf A}_\perp  
= \frac{i}{k^2} {\bf k} \bm{\times} \tilde{\bf B}_\perp. 
\label{Aperp}
\end{eqnarray}
The wave equation (\ref{waveEqsFS}) for $\tilde{\bf B}_\perp$ can then
be written more simply as a wave equation for $\tilde{\bf A}_\perp$:
\begin{eqnarray}
\left( k^2 - \frac{\omega^2}{v^2} \right) \tilde{\bf A}_\perp
&=& \tilde\mu \tilde{\bf J}_\perp.
\label{waveEqsAperp}
\end{eqnarray}
This wave equation depends on the light speed $v$ in the medium, and is 
therefore causal, or more specifically retarded. We shall compare this 
wave equation for $\tilde{\bf A}_\perp$ to that for $\tilde{\bf A}_\parallel$, 
which is still to be found.

To find $\tilde{\bf A}_\parallel$, we begin by putting (\ref{Aperp}) into 
(\ref{curlEq1T}) to get one solution
\begin{eqnarray}
\tilde{\bf E}_\perp 
= i\omega\tilde{\bf A}_\perp 
\label{Eperp}
\end{eqnarray}
out of many mathematically permissible solutions. It would then be natural 
to introduce a $\parallel$ component $\tilde{\bf A}_\parallel$ by the 
equation $\tilde{\bf E}_\parallel = i\omega\tilde{\bf A}_\parallel$, 
so that $\tilde{\bf E} = i\omega\tilde{\bf A}$. 
However, the alternative expression $\tilde{\bf E}_\parallel 
= -ik\tilde{\Phi}$ defined by a mathematically equivalent scalar potential 
$\tilde{\Phi}$ can also be used. 

Historically, different linear combinations of these equivalent forms were 
used that ultimately led to the inclusive expression 
\cite{Jackson01}
\begin{eqnarray}
\tilde{\bf E}_\parallel 
= -ik\tilde{\Phi} + i\omega\tilde{\bf A}_\parallel,
\label{EparallelPhiA}
\end{eqnarray}
and therefore
\begin{eqnarray}
\tilde{\bf E}
= -i{\bf k}\tilde{\Phi} + i\omega\tilde{\bf A}.
\label{EphiA}
\end{eqnarray}

It is clear, however, that (\ref{EparallelPhiA}) contains too much 
freedom of choice. 
The choice $\tilde{\bf A}_\parallel = 0$, called the Coulomb gauge, has the 
advantage that the associated scalar potential called the Coulomb potential
is, like $\tilde{\bf E}_\parallel$ itself, noncausal and instantaneous, 
acting at a distance.

A more general choice called the velocity gauge \cite{Jackson02,Yang05} 
is defined by the {\it gauge condition}
\begin{eqnarray}
\tilde{A}^{(vg)}_\parallel = \alpha\frac{\omega}{k} \tilde{\Phi}^{(vg)},
\label{gaugeCond}
\end{eqnarray}
where the gauge parameter
\begin{eqnarray}
\alpha = \frac{1}{v_g^2}
\label{gaugeAlpha}
\end{eqnarray}
can be written in term of a gauge velocity $v_g$. Putting (\ref{gaugeCond}) 
into (\ref{EparallelPhiA}) yields 
\begin{eqnarray}
\tilde{E}_\parallel 
= -\frac{i}{k} \left( k^2 - \frac{\omega^2}{v_g^2} \right)\tilde{\Phi}^{(vg)}.
\label{EparallelPhi}
\end{eqnarray}
Finally, the Gauss law (\ref{Eparallel}) can be used to eliminate 
$\tilde{E}_\parallel = -i \tilde{\rho}/k\tilde\epsilon_\parallel$ in 
favor of $\tilde{\rho}$ to give $\tilde{\Phi}^{(vg)}$ and 
$\tilde{A}^{(vg)}_\parallel$ as solutions of inhomogenious wave equations
\begin{eqnarray}
\tilde{\Phi}^{(vg)} 
= \frac{\tilde{\rho}}{\tilde{\epsilon}_\parallel (k^2 - \omega^2 /v_g^2)},
\qquad 
\tilde{A}^{(vg)}_\parallel 
= \frac{\omega}{kv_g^2}
\frac{\tilde{\rho}}{\tilde{\epsilon}_\parallel (k^2 - \omega^2 /v_g^2)}.
\label{PhiAparaRho}
\end{eqnarray}
We see that in general, both $\tilde{\Phi}^{(vg)}$ and 
$\tilde{A}^{(vg)}_\parallel$ are causal, with the signal propagated between 
source and field positions at gauge velocity. Since the gauge velocity is 
an arbitrary parameter, this gauge causality is not a physical or 
gauge-independent property. 

It is even more instructive to use (\ref{PhiAparaRho}) in 
(\ref{EparallelPhiA}) to give an expression 
\begin{eqnarray}
\tilde{E}_\parallel 
= \frac{k^2}{k^2 - \omega^2/v_g^2}\tilde{E}_\parallel 
- \frac{\omega^2/v_g^2}{k^2 - \omega^2/v_g^2}\tilde{E}_\parallel,
\label{EparallelPhi+A}
\end{eqnarray}
that shows explicitly that the fractional shares residing in the 
$\tilde{\Phi}^{(vg)}$ and $\tilde{\bf A}^{(vg)}_\parallel$ terms vary with 
the arbitrary gauge velocity. There are thus infinitely many ways to cut 
the $\tilde{E}_\parallel$ cake, but it is always the same cake after all. 
The relation between $\tilde{\Phi}^{(vg)}, \tilde{\bf A}^{(vg)}_\parallel$ 
and $\tilde{\bf E}_\parallel$ is thus not one-to-one, but many-to-one. 
This is the redundant gauge degree of freedom. 

Finally, the Lorenz gauge is realized by the choice 
$v_g = v = 1/\sqrt{\tilde\mu\tilde\epsilon_\perp}$, with the gauge velocity 
set to the physical wave velocity $v$ of the transverse fields. Using this 
physical wave velocity as the gauge velocity does not make the resulting
gauge causality any more physical than that in other choices of $v_g$ for 
the instantaneous $\tilde{E}_\parallel$ field. The nonphysical nature of
gauge causality, including that in the Lorenz gauge, is not sufficiently 
emphasized in many textbooks on electromagnetism.

Another noteworthy feature of the solution $\tilde{A}_\parallel$ given in 
(\ref{PhiAparaRho}) is that it satisfies the wave equation
\begin{eqnarray}
\left( k^2 - \frac{\omega^2}{v_g^2} \right) \tilde{A}^{(vg)}_\parallel
&=& \frac{1}{v_g^2\tilde\epsilon_\parallel} \tilde{J}_\parallel,
\label{waveEqApara}
\end{eqnarray}
where $\omega\tilde{\rho}/k$ has been replaced by $\tilde{J}_\parallel$ 
with the help of the continuity equation (\ref{continuityEq}). This wave 
equation is different from the wave equation (\ref{waveEqsAperp}) satisfied 
by $\tilde{A}_\perp$ unless two conditions are satisfied: 
\begin{eqnarray}
v_g = v, \quad {\rm and} \quad 
\tilde\epsilon_\parallel = \tilde\epsilon_\perp.
\label{2conditions}
\end{eqnarray}
The Lorenz gauge in vacuum is so popular because these conditons are 
satisfied ($v_g = v = c$, and $\tilde\epsilon_\parallel = \tilde\epsilon_\perp 
= \epsilon_0$). Then all three components of $\tilde{A}$ satisfy the same
wave equation. In linear isotropic materials such as plasmas, one finds that 
$\tilde\epsilon_\parallel \neq \tilde\epsilon_\perp$ in general, however. 
So the simplicity of the vacuum Lorenz gauge cannot be maintained for 
many linear materials.

\section{Comparison with Brill and Goodman, and with Yang}
\label{sect:Comparison}

The technique used in this paper depends on L/T separation, and is therefore 
the same technique as that used by Brill and Goodman \cite{Brill67} and by 
Yang \cite{Yang05}. The only difference is that we work in the Fourier 
space where all expressions are algebraic, and therefor much more transparent. 
The resulting simplicity is particularly striking in our discussion of the 
gauge degree of freedom where it is obvious that ${\bf A}_\perp$ is not 
involved at all.  We are also able to 
extend the treatment to linear isotropic materials, a subject of interest 
in materials physics. For electromagnetism in free space already discussed 
in \cite{Brill67,Yang05}, the important gauge independence of the 
electromagnetic fields {\bf E, D, B} and {\bf H} is correctly obtained by 
them, by us, and by everybody else.

There are significant technical differences in the results for the Coulomb 
gauge. In \cite{Brill67} (p. 833), an incorrect Coulomb-gauge condition
$\bm{\nabla \cdot} {\bf A}_\perp = 0$ is used. A nonzero term 
${\bf A}^{(C,\Phi)}$ appears in \cite{Yang05} ((3.30) on p. 745) when it should 
be zero because of the Coulomb-gauge condition $\bm{\nabla \cdot} {\bf A} = 0$.
These problems do not seem to affect the rest of the analyses in
\cite{Brill67,Yang05}.

We differ from these authors in the conclusion drawn from the similar 
technical analyses, however. Our conclusion in its simplest form is that 
the electrostatic potential of a stationary charge in vacuum is just the 
instantaneous Coulomb potential. So the electric field {\bf E} generated 
from it is also instantaneous, acting at a distance. The current {\bf J} 
of this stationary charge is zero. So the charge's vector potential {\bf A} 
and its magnetic induction {\bf B} both vanish. Here then is a simple 
example where an instantaneous interaction is unavoidable. Such a 
literal interpretation of the Maxwell equations in vacuum is accepted 
by many people, including Dirac \cite{Dirac78}. Let us elaborate on our
interpretation.

With the advent of quantum mechanics, the explanation of the instantaneous
Coulomb interaction between two stationary charges in vacuum becomes more 
detailed. In this more modern picture,
the Coulomb interaction arises from the exchange between the interacting 
charges of a virtual photon, the quantum of electromagnetic waves. This
virtual photon is actually a combination of a time-like virtual photon
and a longitudinal virtual photon \cite{Sakurai67}, conforming to the 
requirement of Maxwellian electromagnetism, because these photons are
respectively the quanta of the scalar and longitudinal vector potentials.

A virtual photon in quantum mechanics, like other virtual particles, does
not conserve energy. It can exist momentarily only because of the Heisenberg 
uncertainty principle applied to the energy-time pair of quantized 
dynamical variables. Thus virtual particles do not satisfy the 
energy-momentum relation of real particles ``on the energy shell''. 
In other words, virtual particles are ``off the energy shell''. This 
virtuality is why the potential or field can exist in all space 
around the source, instantaneously. Only real particles ``on the energy 
shell'' that move with a speed $v \leq c$ can carry observable messages or 
signals. Virtual particles cannot. In particular, they cannot be detected 
as energy-violating objects, otherwise energy nonconservation would be 
observed. In other words, virtual particles are given the liberty of 
violating both energy conservation and causality in exchange for being
forever unobservable directly. 

In other respects, real and virtual particles are treated alike. For 
example, the same mathematical technique for describing particle emission 
and absorption is used in quantum mechanics for both real and virtual 
particles, and for both neutral or charged particles. For charged 
particles, Dirac has explicitly written \cite{Dirac78}: 
``Whenever an electron is emitted, 
the Coulomb field around it is simultaneously emitted, forming a kind of 
{\it dressing} for the electron. Similarly, when an electron is absorbed, 
the Coulomb field around it is simultaneously absorbed.'' These processes 
are caused, for example, by the Hamiltonian (the quantized energy operator) 
which ``applies just to one instant of time'', the instant of emission or 
absorption.

In quantum field theory, the physical vacuum ceases to be merely empty 
space. It is instead a sea of vacuum fluctuations, pervasive in all space 
and lasting for all times. An individual vacuum fluctuation exists
momentarily, appearing and disappearing as if by magic. Taken together,
vacuum fluctuations form a relatively smooth fabric on which the
instantaneous Coulomb interaction of macroscopic electromagnetism can be 
realized. 

The present-day picture of the physical vacuum is even more sophisticated. 
In the presence of charges, the vacuum acts like ordinary materials in 
that its permittivity $\epsilon_0$ and permeability $\mu_0$ can also change 
for sufficiently large changes in the space-time scale of the physical 
phenomena involved (\cite{Zee03}, pp. 339--340). 
That is, they are also functions 
of $({\bf k}, \omega)$, or equivalently of the energy scale $M$ of the 
phenomena. Experiments have shown that the effective electromagnetic 
interaction strength $e^2/\epsilon_0$, where $e$ is the electronic charge, 
has increased by 7\% from macroscopic physics where $M \approx 0$ to 
phenomena at the mass scale of $M = 100\:$GeV. It is even expected to 
increase by another factor of more than 2 by the time the so-called grand 
unification scale of $10^{16}\:$GeV is reached. This is the energy 
regime where the weak and strong interactions too can be expected to 
have the same effective coupling strength \cite{Wilczek97}.

It thus appears that the Maxwell equations in vacuum do describe charges 
interacting instantaneously via the ${\bf E}_\parallel$ field, a physical 
process that is not subject to the relativistic limitation of all 
measurable speeds to $c$ or less. However, the Maxwellian theory does not 
explain why this classical picture of action at a distance should hold.

In contrast, \cite{Brill67,Yang05} almost forty years apart both conclude
that the entire ${\bf E}({\bf r},t)$ in vacuum, including  
${\bf E}_\parallel({\bf r},t)$, is retarded. Their conclusion is based on
the assumption that since each of the four contributions (from the three 
vector potential components and one scalar potential) to the Lorenz-gauge 
${\bf E}({\bf r},t)$ is individually retarded, their total 
${\bf E}({\bf r},t)$ must be retarded as well. This assumption is incorrect
because, as we have shown in section \ref{sect:SVpotentials}, the two 
retarded contributions from $\tilde\Phi^{(vg)}$ and $\tilde A^{(vg)}_\parallel$ 
actually add up to an instantaneous $\tilde{\bf E}_\parallel$ for any gauge.  
The gauge-independence of this instantaneous field component is an important
feature of Maxwellian electromagnetism.

\section{The instantaneous and gauge-independent ${\bf E}_\parallel$}
\label{sect:Eparallel}

The unexpected metamorphosis of two gauge causal terms in vacuum into an 
instantaneous interaction in $\tilde{\bf E}_\parallel$ is sufficiently 
important in teaching electromagnetism to merit an explicit demonstration 
of how it materializes in space-time itself. We begin by relating the H
elmholtz and Poisson equations in space
\begin{eqnarray}
\left( \bm{\nabla}^2 + \frac{\omega^2}{v_g^2} \right) 
\frac{e^{\pm i\omega R/v_g}}{4\pi R} 
= - \delta^3({\bf R}) = \bm{\nabla}^2  
\frac{1}{4\pi R}, 
\label{HelmholtzPoisson}
\end{eqnarray}
where ${\bf R} = {\bf r} - {\bf r}'$. The Helmholtz solutions are the 
retarded (upper + sign) and advanced (lower - sign) Green functions of
Jackson \cite{Jackson99}, (6.40) on p. 244. The Poisson Green function
is just the special case when $\omega/v_g = 0$. It is neither retarded nor 
advanced, but instantaneous, as in electrostatics.

In the Fourier space {\bf k}, (\ref{HelmholtzPoisson}) becomes the trivial 
algebraic identity (writing the Poisson side first)
\begin{eqnarray}
k^2 \frac{1}{k^2} = 1 = \left( k^2 - \frac{\omega^2}{v_g^2} \right)
\frac{1}{k^2 - \omega^2/v_g^2}.
\label{HelmholtzPoisson-kspace}
\end{eqnarray}
Left dividing by $\epsilon_0k^2$, right multiplying by the charge density 
$\tilde{\rho}({\bf k},\omega)$, and replacing the latter by the longitudinal
current density $\tilde{J}_\parallel$ from the continuity equation 
(\ref{continuityEq}) in the second term on the Helmholtz side all in one 
step, we get 
\begin{eqnarray}
\frac{1}{\epsilon_0 k^2}\tilde{\rho} = 
\frac{1}{\epsilon_0 k^2}\left( k^2\tilde{\rho} 
- \frac{\omega k}{v_g^2}\tilde{J}_\parallel \right)
\frac{1}{k^2 - \omega^2/v_g^2}.
\label{HPkspace}
\end{eqnarray}
A final left multiplication by $-ik$ gives  
\begin{eqnarray}
\tilde{E}_\parallel({\bf k},\omega) &=& \tilde{E}^{(C)}_\parallel({\bf k},\omega)
= - \frac{ik}{\epsilon_0 k^2}\tilde{\rho} 
\nonumber \\
&=& -ik \tilde{\Phi}^{(vg)} + i\omega \tilde{A}^{(vg)}_\parallel,
\label{Eparakspace}
\end{eqnarray}
where
\begin{eqnarray}
\tilde{\Phi}^{(vg)} &=& \frac{\tilde{\rho}}{\epsilon_0 (k^2 - \omega^2/v_g^2)},
\nonumber \\
\tilde{A}^{(vg)}_\parallel &=& 
\frac{\tilde{J}_\parallel/v_g^2}{\epsilon_0 (k^2 - \omega^2/v_g^2)}.
\label{PhiApara}
\end{eqnarray}
We are now ready to invert the Fourier transform in {\bf k}, as defined by
\begin{eqnarray}
f({\bf r}) = {\cal F}^{-1}\{\tilde{f}({\bf k}) \}
= \int \frac{d^3k}{(2\pi )^3} e^{i{\bf k}\bm{\cdot}{\bf r}} 
\tilde{f}({\bf k}). 
\label{FInversion-k} 
\end{eqnarray}

We shall need the convolution or folding theorem \cite{Wong91} and the 
inverse Fourier transform from (\ref{HelmholtzPoisson})
\begin{eqnarray}
{\cal F}^{-1}\{\tilde{f}({\bf k}) \tilde{g}({\bf k})\}
&=& [f \ast g]({\bf r})
= \int d^3r' f({\bf r}-{\bf r}')g({\bf r}'), \\
\label{FoldingThm} 
{\cal F}^{-1} \left\{ \frac{1}{k^2 -\omega^2/v_g^2} \right\} 
&=& \frac{e^{\pm i\omega r/v_g}}{4\pi r}.
\label{HelmholtzGF}
\end{eqnarray}
A straightforward calculation then yield the desired connection between the 
results in the instantaneous Coulomb gauge and the general velocity gauge 
in the space $({\bf r}, \omega/t)$, where $\omega/t$ means either 
$\omega$ or $t$:
\begin{eqnarray}
{\bf E}_\parallel^{(C)}({\bf r}, \omega/t) = 
- \bm{\nabla}\Phi^{(vg)}({\bf r}, \omega/t) 
+ i\omega{\bf A}^{(vg)}_\parallel({\bf r}, \omega/t),
\label{EparaC+vGauge}
\end{eqnarray}
where the $i\omega$ factor in the second term on the velocity-gauge side 
stands for the time differential operator $-\partial_t$ if the space is
$({\bf r},t)$. (\ref{EparaC+vGauge}) is the desired instantaneous and 
gauge-independent longitudinal electric field.

The functions in the $({\bf r},\omega)$ space that appear in 
(\ref{EparaC+vGauge}) can be obtained by straightforward (inverse) Fourier 
transformations: 
\begin{eqnarray}
\tilde{\bf E}^{(C)}_\parallel({\bf r}, \omega) &=& 
-\bm{\nabla} \int d^3{r}' \frac{\tilde{\rho}({\bf r}', \omega)}
{4\pi\epsilon_0 |{\bf r} - {\bf r}'|}; \nonumber \\
\tilde{\Phi}^{(vg)}_\parallel({\bf r}, \omega) &=& \int d^3{r}' 
\frac{e^{\pm i\omega |{\bf r} - {\bf r}'|/v_g} \tilde{\rho}({\bf r}', \omega)}
{4\pi\epsilon_0 |{\bf r} - {\bf r}'|}, \nonumber \\
\tilde{\bf A}^{(vg)}_\parallel({\bf r}, \omega) &=& 
 \int d^3{r}' \frac{e^{\pm i\omega |{\bf r} - {\bf r}'|/v_g} 
\tilde{\bf J}_\parallel({\bf r}', \omega)/v_g^2}
{4\pi\epsilon_0 |{\bf r} - {\bf r}'|},
\label{Epara-rw}
\end{eqnarray}
where $\tilde{\bf J}_\parallel$ is from the Helmoltz theorem 
(\ref{HelmholtzThm3}) in space
\begin{eqnarray}
\tilde{\bf J}_\parallel({\bf r}', \omega) &=& -\bm{\nabla}' \int d^3r''
\frac{\bm{\nabla}'' \bm{\cdot} \tilde{\bf J}_\parallel({\bf r}'',\omega)}
{4\pi|{\bf r}' - {\bf r}''|}  
\nonumber \\
&=& -i\omega \bm{\nabla}' \int d^3r''
\frac{\tilde{\rho}({\bf r}'', \omega)}{4\pi|{\bf r}' - {\bf r}''|}. 
\label{Apara-rw}
\end{eqnarray}
The corresponding functions in $({\bf r}, t)$ are easily obtained:
\begin{eqnarray}
{\bf E}^{(C)}_\parallel({\bf r}, t) &=& 
-\bm{\nabla} \int d^3{r}' \frac{{\rho}({\bf r}', t)}
{4\pi\epsilon_0 |{\bf r} - {\bf r}'|}; \nonumber \\
{\Phi}^{(vg)}_\parallel({\bf r}, t) &=& \int d^3{r}' 
\frac{{\rho}({\bf r}', t \mp |{\bf r} - {\bf r}'|/v_g)}
{4\pi\epsilon_0 |{\bf r} - {\bf r}'|}, \nonumber \\
{\bf A}^{(vg)}_\parallel({\bf r}, t) &=&  
\int d^3{r}' 
\frac{{\bf J}_\parallel({\bf r}', t \mp| {\bf r} - {\bf r}'|/v_g)/v_g^2}
{4\pi\epsilon_0 |{\bf r} - {\bf r}'|},\nonumber \\
{\bf J}_\parallel({\bf r}', t_r) &=&
\partial_t \bm{\nabla}' \int d^3r''
\frac{{\rho}({\bf r}'', t_r)}{4\pi|{\bf r}' - {\bf r}''|}. 
\label{EparaPhiA-rt}
\end{eqnarray}
They agree with those found in \cite{Brill67,Yang05}

From the perspective of teaching electromagnetism at an introductory level,
the solutions of the Helmholtz equation given in (\ref{HelmholtzPoisson})
may be considered too advanced a result. The folding theorem 
(\ref{FoldingThm}) is quite elementary, however, once Fourier transforms 
are used. The folding theorem alone enables the concepts of locality in 
space and instantaneity in time to be discussed at an introductory
qualitative level, without the need to actually work out the precise
inverse transforms. This is possible because the folding theorem shows
that two functions with the same {\bf k} dependence must be localized
relative to each other in space. This is so because both functions are 
the same function in space. Similarly, two functions with the same $\omega$ 
dependence must be instantaneous in time. Conversely, two functions
with different {\bf k} (or $\omega$) dependences must be nonlocal 
(or non-instantaneous) in space (or time). 

Let us then use this qualitative instantaneity/localization test on 
certain $\omega$/{\bf k} dependent expressions we have constructed in the 
preceding sections. Consider first electromagnetism in free space, where  
$\epsilon_0$ and $\mu_0$ are independent of {\bf k} or $\omega$ in the 
Maxwellian theory. Then  $\tilde{E}_\parallel({\bf k}, \omega) = 
-i\tilde{\rho}({\bf k}, \omega)/k\epsilon_0$ of (\ref{Eparallel}) and 
$\tilde{\rho}({\bf k}, \omega)$ have the same $\omega$ dependence but 
different {\bf k} dependences. Hence
${\bf E}_\parallel({\bf r}, t)$ is instantaneous with, but not localized 
at, the charge density $\rho({\bf r}, t)$. That is, 
${\bf E}_\parallel({\bf r}, t)$ is action at a distance. By the same 
test, both the transverse fields $\tilde{\bf E}_\perp({\bf k}, \omega)$ 
and $\tilde{B}_\perp({\bf k}, \omega)$ are non-instantaneous (i.e., 
causal) and nonlocal in space-time.
 
In linear materials, where $\tilde\epsilon$ and $\tilde\mu$ depend on 
{\bf k} and $\omega$, the test shows that in 
$\tilde{E}_\parallel= -i\tilde{\rho}/k\tilde\epsilon_\parallel$, the
instantaneity between $E_\parallel$ and $\rho$ is lost. However, there is 
still instantaneity between $E_\parallel$ and the polarization charge 
contained in $\tilde\rho/\tilde\epsilon_\parallel$. That is, the polarized
response of the material is retarded relative to the external $\rho$, but
as the polarization charges appear, they generate their electric fields
simultaneously and instantaneously, in exactly the way Dirac has described.

\section{Gauge transformations}
\label{sect:GaugeX}
 
Even after a gauge is chosen, the resulting potentials are still not 
uniquely determined, as we shall now demonstrate.

Consider the general gauge change or transformation
\begin{eqnarray}
\tilde{\bf A}_\parallel &\rightarrow& 
\tilde{\bf A}'_\parallel 
= \tilde{\bf A}_\parallel + i{\bf k}\tilde{\chi},
\nonumber \\
\tilde{\Phi} &\rightarrow& 
\tilde{\Phi}' = \tilde{\Phi} + i\omega\tilde{\chi},
\nonumber \\
\alpha &\rightarrow& \alpha' = \alpha + \beta.
\label{gaugeX}
\end{eqnarray}
The change $\Delta\tilde{\bf A}_\parallel = ik\tilde{\chi}$ comes from the
spatial vector field $\bm{\nabla}\chi({\bf r},t)$ generated from a scalar
field $\chi({\bf r},t)$ whose dimensional unit of measurement is chosen to 
be consistent with the dimension of ${\bf A}({\bf r}, t)$. The change 
$\Delta\tilde{\Phi} = i\omega\tilde{\chi}$ then has the same dimension 
as $\tilde{\Phi}$. Furthermore, the resulting electric field component
\begin{eqnarray}
\tilde{\bf E}'_\parallel 
= -ik\tilde{\Phi}' + i\omega\tilde{\bf A}'_\parallel
= \tilde{\bf E}_\parallel
\label{EparallelPhiAnew}
\end{eqnarray}
remains unchanged, i.e., gauge-independent, for any gauge function 
$\tilde{\chi}$.

We should now mention the indispensable role $A_\parallel$ plays indirectly 
in the Lorentz covariance of the Maxwell equations discovered by Einstein in 
1905 \cite{Einstein05}. This Lorentz covariance refers to the invariance in 
form of the Maxwell equations under a Lorentz transformation to another 
special inertial frame called a Lorentz frame (where the time variable also 
changes in order to keep light speed at its universal value $c$). One can find 
in most textbooks on electromagnetism how this Lorentz covariance is assured 
if the 4-potential $\tilde{A}_\alpha = (\tilde{\bf A}, i\tilde{\Phi}/c)$ 
is a 4-vector in the Minkowski 4-dimensional spacetime. It then transforms 
under Lorentz transformations like any other 4-vector such as 
$k_\alpha = ({\bf k}, i\omega/c)$. $\tilde{A}_\parallel$ is a needed component 
of this 4-potential. Furthermore, the new potential defined by (\ref{gaugeX}), 
namely $\tilde{A}'_\alpha = \tilde A_\alpha + i\tilde{\chi} k_\alpha$ 
is also a 4-vector, transforming correctly under Lorentz transformations. 
Electrodynamics, including quantum electrodynamics, becomes mathematically
simpler and physically more transparent when expressed in terms of these 
4-potentials.

The yet undetermined gauge function $\tilde{\chi}$ must also satisfy the 
gauge condition
\begin{eqnarray}
\tilde{A}'_\parallel = (\alpha + \beta) \frac{\omega}{k} \tilde{\Phi}'.
\label{gaugeCondnew}
\end{eqnarray}
Used with the new potentials, this equation can be simplified into 
the inhomogeneous wave equation 
\begin{eqnarray}
\left[ k^2 - (\alpha + \beta)\omega^2 \right] \tilde{\chi}
&=& -i\omega\beta \tilde{\Phi}.
\label{waveEqChi}
\end{eqnarray}
For $\beta = 0$, the gauge condition or choice (\ref{gaugeCond}) is 
unchanged. The driving term on the right of (\ref{waveEqChi}) then 
vanishes. The resulting homogeneous wave
equation is satisfied either when $\tilde{\chi} = 0$ (an uninteresting 
possibility or a trivial solution), or when $\tilde{\chi} = \tilde{\chi}_0 
\neq 0$, but $k^2 = (\alpha + \beta)\omega^2$. Since $k^2$ comes from 
the space differential operator $\nabla^2$, while $\omega^2$ comes from 
the time differential operator $\partial_t^2$, the second possibility is 
satisfied only by the solutions $\chi_0({\bf r},t)$ of the homogeneous wave 
equation in space-time. The functions $\tilde{\chi}_0({\bf k}, \omega)$ 
that now appear in (\ref{waveEqChi}) are their Fourier transforms. These 
gauge functions define a class of gauge changes called restricted gauge 
transformations. 

The gauge change (\ref{gaugeX}) even allows for a change of the gauge 
condition if $\beta \neq 0$. Then an additional gauge function that is the 
solution of the inhomogeneous wave equation (\ref{waveEqChi}), namely
\begin{eqnarray}
\tilde{\chi} = -i \frac{\omega\beta \tilde{\Phi}}
{k^2 - (\alpha + \beta)\omega^2}, 
\label{chi}
\end{eqnarray}
is needed in addition to those for restricted gauge transformations that 
do not change the gauge condition. These more general gauge transformations 
are called unrestricted gauge transformations.

Other gauges have been used with gauge conditions specified in space-time. 
For such gauges, our treatment in the Fourier space $({\bf k},\omega)$ does 
not appear to offer any advantage. These special gauges have been discussed 
in the recent review by Jackson \cite{Jackson02}.

\section{The physical origin of the gauge degree of freedom}
\label{sect:GaugeOrigin}

Why should there be a gauge degree of freedom? We follow the explanation 
given by Zee (\cite{Zee03}, pp. 168, 171, 456), and elaborate on it.

The physical origin of the gauge degree of freedom lies at one of the most 
striking features of the Maxwell equations, a feature abstracted by Einstein 
into one of the two pillars of his special theory of relativity. As he 
wrote in his epochal 1905 paper on ``The electrodynamics of moving bodies'' 
\cite{Einstein05}, ``light is always propagated in empty space with a 
definite velocity $c$ which is independent of the state of motion of the 
emitting body.'' In other words, the direction of light propagation in 
vacuum might change in a different Lorentz frame of reference. Light 
speed in vacuum will never decrease, and certainly will not decrease to zero.
Hence in every Lorentz frame where a light beam appears, there is a unique 
direction $\hat{\bf e}_{\bf k} = \hat{\bf e}_\parallel$ along which it 
propagates. Each of the electromagnetic fields ${\bf E}_\perp, {\bf B}_\perp$ 
responsible for its propagation is always made of two independent components 
in the perpendicular directions ${\bf e}_1$ and ${\bf e}_2$ in that Lorentz
frame. As we say in optics, light has only two directions of polarization.
We shall explain below how these two polarization states materialize in 
quantum mechanics as the two transverse states of the intrinsic spin of the 
photon, the quantum of electromagnetic waves. 

When relativity is applied to the kinematics of a point particle of mass $m$, 
velocity {\bf v}, momentum {\bf p}, and energy $E$ in a Lorentz frame, 
Einstein obtained the famous energy-momentum relation
\begin{eqnarray}
E^2 - {\bf p}^2 = m^2,
\label{RelEMom}
\end{eqnarray}
for the relativistic momentum
\begin{eqnarray}
{\bf p} = \gamma m {\bf v}, \quad {\rm where} \quad 
\gamma = \frac{1}{\sqrt{1 - v^2/c^2}}.
\label{RelMom}
\end{eqnarray}
This relativistic kinematics restricts the velocity $v$ of all massive 
particles of finite momentum or energy to finite values of $\gamma$, 
and to velocities less than $c$:
\begin{eqnarray}
v = c \sqrt{1- \gamma^{-2}} < c.
\label{VelocityBound}
\end{eqnarray}

If light is considered a particle, its invariant velocity $c$ will 
require that $\gamma = \infty$. When Compton scattering was discovered 
\cite{Compton23}, light was found to have indeed the attributes of a 
particle of finite energy $E$ and momentum {\bf p}. These attributes are 
consistent with (\ref{RelMom}) only if the mass of a photon is exactly zero.

A photon in a beam of light described by the Fourier basis function 
$\psi_{{\bf k}, \omega}({\bf r}, t)$ of (\ref{FourierBasis}) has a wave 
vector {\bf k} and frequency $\omega$. The quantum wave-particle duality 
of the photon is then expressed by the quantum relations
\begin{eqnarray}
{\bf p} = \hbar{\bf k}, \quad {\rm and} \quad E = \hbar\omega,
\label{quantumEp}
\end{eqnarray} 
where $\hbar$ is the reduced Plank constant. The photon, like any other
particle, also has an orbital angular momentum vector $\bm{\ell} = 
{\bf r} \times {\bf p}$. This vector must therefore lie on the 12 plane 
perpendicular to ${\bf e}_{\bf k} = {\bf e}_3$, with only two independent 
components $\ell_1$ and $\ell_2$, just like ${\bf A}_\perp$, in any 
Lorentz frame. We thus see that the absence of a physically meaningful 
and therefore gauge-independent $A_\parallel$ component has the same 
physical origin as the absence for the massless photon of the component 
$\ell_3 = \ell_\parallel$ of its orbital angular momentum vector, namely
the fact that the massless photon always moves with light speed $c$ in 
vacuum in any Lorentz frame. 

The inaccessibility of the longitudinal components of these vectors {\bf A} 
and $\bm{\ell}$ may appear a little unsatisfactory, but modern physics, 
i.e., special relativity and quantum mechanics, now come to the 
rescue. Consider first a massive particle that can only move with speeds
$v < c$. It is then possible to go to a Lorentz frame where the particle
is at rest. In this rest frame, the particle's orbital angular momentum 
vanishes because its momentum {\bf p} vanishes. Now it has been found 
in quantum mechanics that all massive particles at rest have an intrinsic 
spin or intrinsic angular momentum vector  {\bf s} whose length can only 
take one of the permissible values of $s=0, 1/2, 1, 3/2, ...$ (in units 
of $\hbar$). In addition, its projection $s_z$ along any coordinate axis 
${\bf e}_z$ in space can only have values that differ from one another 
by an integer (in units of $\hbar$). Since $s_z$ is a projection of 
{\bf s}, it also satisfies the constraint $|s_z| \leq s$. Both constraints 
are satisfied by the $2s + 1$ possible values of $s_z$ in the range 
$-s \leq s_z \leq s$. For example, 
a particle with $s = 1$ can only have the three $z$-projections of 
$s_z = 1, 0,$ or $-1$. Furthermore, in the rest frame of a massive particle, 
there is no preferred direction in space. All three components of its 
intrinsic spin vector {\bf s} are dynamically equivalent.

The spin vector {\bf s} is also called an angular momentum vector, because 
it is found to have all the properties of the particle's orbital angular 
momentum vector $\bm{\ell}$ 
when the particle is moving. Only then can the two angular momenta of a 
moving particle be added together to form a total angular momentum 
${\bf j} = \bm{\ell} + {\bf s}$, as Pauli \cite{Pauli25}, and Uhlenbeck 
and Goudsmit \cite{Uhlenbeck25} found in their explanation of the 
anomalous Zeeman effect. (The anomalous Zeeman effect has its origin in
the unexpected doubling of the number of atomic energy levels in certain 
atoms in a magnetic field. It is caused by the intrinsic spin of the 
electron.) So for massive particles in motion, all three components of 
their orbital and total angular momenta are dynamically equivalent as well.

Massive particles of intrinsic spin $s = 1$ are called vector bosons, 
vector because its {\bf s} vector has three quantized $s_z$ values, 
matching the number of components of a vector in space. Any particle 
whose intrinsic spin $s$ is an integer is called a boson, named after 
Bose, the famous Indian contemporary of Einstein. The photon too has 
been found to be a vector boson. If the photon were massive, it 
too would have a rest frame where there is no preferred direction in 
space. Its three intrinsic spin projections $s_z$ would be just the 
three polarization states. Then all three components of its vector 
potential {\bf A} would 
be accessible and dynamically equivalent, and the gauge redundancy 
would disappear. Quantum mechanics would thus work its magic in the 
massive photon's rest frame, turning the ugly toad of gauge redundancy 
into a handsome prince of dynamical wholesomeness.

So, as a massive photon becomes massless, it acquires the universal light 
speed $c$ in all Lorentz frames. It loses its rest frame. Condemned to 
perpetual motion, both its orbital and intrinsic angular momentum vectors
become forever confined to the transverse plane perpendicular to its 
direction of motion. Its $\tilde{A}_\parallel$ too decouples from the 
two dynamical transverse components. It is made accessible to dynamics 
only by sharing the function with the scalar potential. It becomes our 
redundant gauge degree of freedom.

\appendix

\section{Helmholtz theorem}
\label{sect:HelmholtzThm}

The usual and customary integral form of the Helmholtz decomposition of a 
vector field in space into a gradient part and a curl part can be obtained 
by evaluating the inverse Fourier transform of (\ref{HelmholtzThm}). 
The $\omega$ variable is actually not involved, and will be suppressed here. 
Hence we consider the inverse Fourier transform ${\cal F}^{-1}$, as defined 
by (\ref{FR-E}) and reproduced here without its $\omega$ or $t$ dependence as 
\begin{eqnarray}
{\bf E}({\bf r}) = {\cal F}^{-1}\{\tilde{\bf E}({\bf k}) \}
= \int \frac{d^3k}{(2\pi )^3} e^{i{\bf k}\bm{\cdot}{\bf r}} 
\tilde{\bf E}({\bf k}). 
\label{FInversion} 
\end{eqnarray}
Such inverse Fourier transforms can handle the unit vector ${\bf e}_{\bf k}$ 
readily only in the original vector/scalar form {\bf k}/$k$. So we begin
by writing the vector field $\tilde{\bf E}({\bf k})$ in (\ref{HelmholtzThm})
in the alternative dyadic form
\begin{eqnarray}
\tilde{\bf E}({\bf k}) = \tilde{\bf E}_\parallel + \tilde{\bf E}_\perp
= {\bf k} \frac{1}{k^2} ({\bf k} \bm{\cdot} \tilde{\bf E}) 
- {\bf k} \bm{\times} \frac{1}{k^2}\;\, ({\bf k} \bm{\times} \tilde{\bf E}). 
\label{HelmholtzThm2}
\end{eqnarray}

The inverse Fourier transform {\bf E}({\bf r}) of (\ref{HelmholtzThm2}) can 
be found readily by using inverse transforms such as
\begin{eqnarray}
{\cal F}^{-1}\{ {\bf k} \} = -i \bm{\nabla}, \\
{\cal F}^{-1}\{ i{\bf k}\bm{\cdot}\tilde{\bf E} \} 
=  \bm{\nabla \cdot} {\bf E}({\bf r}), \\
{\cal F}^{-1} \left\{ \frac{1}{k^2} \right\} = \frac{1}{4\pi |{\bf r}|}.
\label{CoulombGF}
\end{eqnarray}
Each term on the right in (\ref{HelmholtzThm2}) contains three {\bf k} 
factors. The middle $1/k^2$ factor can be treated in two equivalent ways
giving the same nonlocal kernel. The nonlocal kernel comes either from
the convolution or folding theorem (\ref{FoldingThm}) applied to a product 
of functions of $k$ in the Fourier space {\bf k}, or from the nonlocal 
operator $1/k^2$ in space that requires an integral over {\bf r}$'$, 
as done in quantum mechanics. 
In both cases, one finds the integral Helmholtz theorem in space, with 
\begin{eqnarray}
{\bf E}_\parallel({\bf r}) = -\bm{\nabla}
\int d^3r'\frac{1}{4\pi|{\bf r} - {\bf r'}|} \bm{\nabla}' \bm{\cdot} 
{\bf E}_\parallel({\bf r}'), 
\nonumber \\
{\bf E}_\perp({\bf r}) = \bm{\nabla} \times 
\int d^3r'\frac{1}{4\pi|{\bf r} - {\bf r'}|} \bm{\nabla}' \times 
{\bf E}_\perp({\bf r}'). 
\label{HelmholtzThm3}
\end{eqnarray}

\end{document}